# ARM 7 Based Controller Area Network for Accident Avoidance in Automobiles

Kashyap Joshi[#1], Vipul Gohil[*2]

[#] *Student, Electronics and Telecommunication Department, NMIMS University*
*Parle(west), Mumbai, India.*

[*] *Assistant Professor , Electronics and Telecommunication Department, NMIMS University*
*Parle(west), Mumbai, India.*

[1] `kashyapjoshi169@yahoo.co.in`
[2] `vipul.gohil@nmims.edu`

*Abstract*— Based on requirements of modern vehicle, in- vehicle Controller Area Network (CAN) architecture has been implemented. In order to reduce point to point wiring harness in vehicle automation, CAN is suggested as a means for data communication within the vehicle environment. The benefits of CAN bus based network over traditional point to point schemes will offer increased flexibility and expandability for future technology insertions.
  This paper describes system which uses sensors to measure various parameters of the car like speed, distance from the other car, presence of alcohol in car and accidental change of lane and sends a warning signal to the driver if any of the parameter goes out
of range to avoid accidents . In addition to this if accident occurs in any remote area then using bump sensor accident is detected and SMS is send immediately using GSM. A situation that provides a good example of how the system works is when a driver is about to change lanes, and there is a car in his blind spot. The sensors will detect that car and inform the driver before he starts turning, preventing him from potentially getting into a serious accident.

## I. INTRODUCTION

Nowadays accidents occur due to mistakes done by driver. An intelligent system needs to be developed to overcome these mistakes. So this system is proposed where mistakes done by driver are eliminated. Most of the intelligent car systems have monitoring system only. Antilock brakes, speed sensors and other automatic systems are present in sports cars and other luxury cars only. But these cars are not affordable to everyone. So, a system needs to be developed which can be implemented in every car

 A collision avoidance system is a system of sensors that is placed within a car to warn its driver of any dangers that may lie ahead on the road. Some of the dangers that these sensors can pick up on include how close the car is to other cars surrounding it, how much its speed needs to be reduced while going around a curve, and how close the car is to going off the road.

   The system uses sensors that send and receive signals from things like other cars; obstacles in the road, traffic lights, and even a central database are placed within the car and tell it of any weather or traffic precautions. A situation that provides a good example of how the system works is when a driver is about to change lanes, and there is a car in his blind spot. The sensors will detect that car and inform the driver before he starts turning, preventing him from potentially getting into a serious accident.

   Ultrasonic sensor is adapted to measure the distance with respect to the previous car. For rear-end end collision avoidance subsystem, the currently available ultrasonic sensors for vehicles are adopted for approaching cars with relatively low speed. While the rough reading of distance data cannot be applied directly, an intelligent approach is proposed to process the raw distance readout of sensors to produce appropriate warning signals. Also alcoholic sensors is included in the car to monitor the person in the car; if the person appears to be drunk the transmission will be automatically switched off. If accident occurs then bump sensor detects accident and immediately sends SMS to hospitals and police station about location of accident.

## II. OVERVIEW OF CAN PROTOCOL

   Controller area network (CAN) provide high reliability and good real-time performance with very low cost. Due to this, CAN is widely used in a wide range of applications, such as in-vehicle communication, automated manufacturing and distributed process control environments.CAN bus is a serial data communication protocol invented by German BOSCH Corporation in 1983. CAN is a network protocol which is designed for the car industry [1].

   Since data communication in car often have many sensors transmitting small data packets, CAN supports data frames with sizes only up to 8 bytesas shown in Figure 1. Meanwhile, the 8 bytes will not take the bus for a long time, so it ensures real-time communication. CAN use a large amount of overhead, which combined with a 15-bit CRC makes CAN very secure and reliable.

   CAN protocol use non-destructive bitwise arbitration process to access shared resource. CAN protocol define a logic bit 0 as a dominant bit and a logic bit 1 as a recessive bit, each transmitting node monitors the bus state and compares the received bit with the transmitted bit [2]. If a dominant bit is received when a recessive bit is transmitted then the node stops transmitting (i.e. it lost arbitration).

Arbitration is performed during the transmission of the identifier field. There are two message formats: Base frame format with 11 identifier bits and extended frame format with 29 identifier bits [3].

| Start of Frame | Arbitration Field | Control Field | Data Field (up to 8 bytes) | CRC Field | ACK Field | End of Frame |
|---|---|---|---|---|---|---|

**Figure 1**. CAN Data Frame

*A. CAN Bus Electrical Characteristics*

CAN transmission medium formed by the two, One is called high-level transmission line CANH and another is called low-level transmission line CANL, connected to CANH and CANL pins of MCP2551 CAN transceiver. $V_{CANH}$ and $V_{CANL}$ be the voltage level of CANH and CANL lines with respect to ground. The difference between them is called difference voltage $V_{diff}$.

*B. Hierarchical structure of CAN BUS*

Architecture of CAN protocol based on OSI reference model is as shown in Figure 2. CAN protocol contain only three layers, physical layer, data link layer and application layer. Application layer has different protocols such as SAE J1939, CANopen, DeviceNet, etc [4].

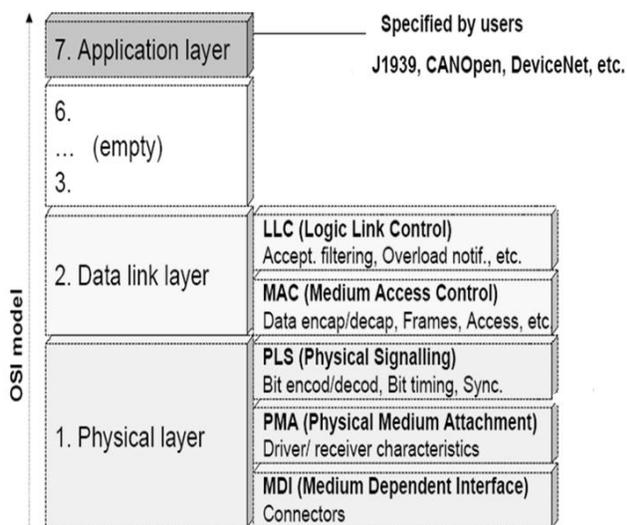

Figure 2. Hierarchical structure of CAN BUS

*C. Scheduling of CAN BUS*

CAN protocol implements fixed priority scheduling of CAN messages. Higher priority node has lower node ID. If available bandwidth is scarce, problems come with traditional fixed-priority based scheduling [5]. It is possible that low priority control loops cannot access the network all the time, since limited resources have been consumed by high priority loops. As a result of tremendous delays, low priority control loops may be destabilized. The problem of fixed priority is overcome using direct feedback scheduling algorithm, namely MUF (maximum urgency first) is integrated in the network scheduler.

Upon invocation, the scheduler calculates the urgency of each control loop based on the set points and current system outputs. According to the MUF algorithm, the scheduler produces new priorities based on these urgency values. And then, messages in different loops will be transmitted in accordance with the newly assigned priorities[6].A new mixed traffic schedule (MTS) is based on the communication principle of controller area network, network scheduling and analysis of schedule.

The core idea of MTS is set the relative deadline information into the identifier. Use the earliest deadline first (EDF) message scheduling algorithm for high priority information and the RMS (rate monotonic scheduling) message scheduling algorithm for low priority information.

*D. Reliability*

Reliability is defined as the probability of no failures in an operational interval. High error handling capability of CAN improves system reliability. If any message transmitting node has detected an error, the node forcibly aborts transmission. Then it attempts to retransmit repeatedly until its message is transmitted successfully. This functionality may let the CAN bus be hogged, if the node of high priority is failed. It is the designer's responsibility to ensure that no any message node hogs the bus. To avoid such crisis, the faculty of the transmit error counter (TEC) and the receive error counter (REC) are started to diagnose the conditions of CAN controller [7]. MCP2515 CAN controller has TEC and REC which enhances reliability of CAN bus system.

A CAN controller can be in one of three states: error active, error passive or bus off state. The operating state of the controller is controlled by two error counters – TEC and REC. The CAN controller is in error active state if TEC less than 127 and REC less than 127 [8]. Passive state is used if (TEC greater than 127 or REC greater than 127) and TEC less than 255. Bus off state is entered if TEC is greater than 255. Once the CAN controller has entered bus off state, it must be reset by the host microcontroller in order to be able to continue operation.

### III. HARDWARE DESIGN

The proposed block diagram for CAN bus communication system is as shown in Figure 3. In this system the ultrasonic sensor is mounted on the front and backside of the car for measuring the distance between the two cars and if the distance is less then to avoid accident warning signal will be given to the driver on the LCD

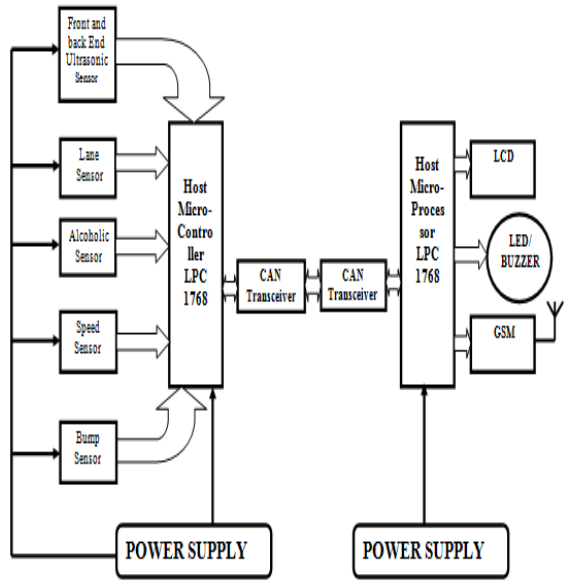

Figure 3 Block Diagram

The alcoholic sensor will sense whether the driver is drunk and if the driver is drunk then the driver will not be allowed to start the car . If the car accidently changes its lane then it will be detected using an IR sensor and buzzer will be turned on.

The speed sensor will monitor the speed of the car and if found high then warning will be given to the driver using an alarm.

Here the sensors will communicate with the output devices using CAN (Control Area Network) protocol which will be implemented in the AVR controller

A. *Ultrasonic sensor*

Ultrasonic sensor is adapted to measure the distance with respect to the previous car. While the car is in motion the distance of another car is measured and accordingly warning signals are given to the driver.

B. *Alcoholic sensor*

Alcoholic sensors in it to monitor the person in the car. If the person appears to be drunk the transmission will be automatically switched off.

C. *Speed sensor*

This sensor monitors the speed of the car and if the speed is found to be more than a prescribed level then a warning signal will be given to the driver.

D. Lane sensor

Lane sensor will detect whether the car is in the same lane it is travelling and if accidently the car changes its lane then acccordingly a warning will be given to the driver.

A dominant state occurs w h e n the differential voltage between CANH and CANL is greater than a defined voltage (e.g., 1.2V). A recessive state occurs when the differential

E. Bump Sensor

The bump sensor detects accidents and if accident is detected then a message is send a message to hospital and police station about location of accident

F. *Microcontroller*

The controller will take input from the sensors and depending on the various sensor inputs output devices will be driven using the other microcontroller. The controller LPC 1768 has inbuilt CAN controller .The two microcontroller will communicate with one another using can communication protocol. LPC 1768 is a Cortex M-3 controllers for embedded applications featuring a high level of integration and low power consumption. The ARM Cortex-M3 is a next generation core that offers system enhancements such as enhanced debug features and a higher level of support block integration.

G. *Host Micro- processor*

The host processor decides what the meaning of received messages is and accordingly takes action on output devices. Output devices are connected to the host processor.

H. *Transceiver*

CAN transceiver MCP2551 adapts signal level from the bus to level that the CAN controller expects and has protective circuitry that protects the CAN controller. It converts the transmit-bit signal received from the CAN controller into a signal that is sent onto the bus.

I. *Output Devices*

According to the inputs from various sensors warning signals are given to the driver to lower the speed of the car or stop turning to get into wrong lane. Also GSM is used to send message in case of accident occurs.

The microcontroller has inbuilt CAN specification, version 2.0B. It is capable of transmitting and receiving both standard and extended data and remote frames with 0 – 8 byte length of the data field. It has two acceptance masks and six acceptance filters that are used to filter out unwanted messages, thereby reducing the host MCUs overhead[9]. It also has three transmit and two receive buffers with prioritized message storage and three transmit buffers with prioritization and abort features.

Typically, each node in a CAN system must have a device to convert the digital signals generated by a CAN controller to signals suitable for transmission over the bus cabling (differential output). It allows a maximum of 112 nodes to be connected and a nominal termination resistor value of 120Ω[9].

voltage is less than a defined voltage (typically 0V). The RXD output pin reflects the differential bus voltage between CANH and CANL[9].

## IV. PROPOSED ALGORITHM

Algorithm for the proposed system is divided in two parts as follows: Transmitter and Receiver

### A. Transmitter

Algorithm for transmitter side which consists sensors, AVR microcontroller and CAN (MCP2515) is as follows:
1. Initialize SPI (Serial Peripheral Interface).
2. Initialize LCD.
3. Initialize CAN (MCP2515).
4. Provide impulse to ultrasonic sensor.
5. Measure distance from other car and display on LCD.
6. Transmit distance via CAN (MCP2515).
7. If alcohol sensed send X else go to step 8.
8. If lane cutting detected send Y else go to step 9.
9. Sense speed and if speed goes beyond range send Z else go to step 10.
10. Check for impact and if impact detected send A else go to step 4.

### B. Receiver

Algorithm for transmitter side which consists output devices, AVR microcontroller and CAN (MCP2515) is as follows:
1. Initialize SPI (Serial Peripheral Interface).
2 . Initialize LCD.
3 . Initialize CAN (MCP2515).
4 . Send acknowledgment to the transmitter.
5. Receive distance data from CAN of transmitter and if distance is less then display warning signal on LCD.
6. If X is received then display "Car cannot be started" else go to step 7.
7. If Y is received then display "Wrong lane" else go to step 8.
8. If Z is received then turn on buzzer else go to step 9.
9. If A is detected then send SMS through GSM else go to step 5.

## V. DESIGN SCHEME OF COMMUNICATION PROTOCOL

The design scheme of communication protocol is explained in this section. Identifier of the message is the unique character for the application program to distinguish messages. In this communication system, when a node receives a message correctly (until the last bit of the EOF area is right), the configured filter box message, and then save the messages with matched ID in receiving box. By using this feature, communication protocol can be made. Different identifiers are set for every data type or control command in this system, then distinguish the received messages conveniently, and choose corresponding processing mode[10].

The standard format of identifier is used in this system as shown in Figure 4[9]. It has 11 bits. Use of standard identifier can reduce the data length and improve data transmission efficiency. In this system, the 11 bit identifier is designed for the "address code + type code" format. Bits D7 to D4 of identifier is the address field, providing at most 16 address codes, and every address code corresponds to a individual node.

Bits D3 to D0 is the type field, which can also provide 16 type codes. And the bits D10 to D8 is the backup filed which is used for system expansion. By configuring the value of the filter ID, each node would only receive the messages with the matched address code[10].

| D10 D9 D8 | D7 | D6 | D5 | D4 | D3 | D2 | D1 | D0 |
|---|---|---|---|---|---|---|---|---|
| 000 | 0 | 0 | 0 | 1 | 0 | 0 | 1 | 0 |
| backup | address code | | | | type code | | | |

Figure 4. Identifier format[9]

## VI. CAN BUS NETWORK SOFTWARE DESIGN

The CAN version 2.0B has three transmit buffer and two receive buffers. The CAN message transmission flow is as shown in Figure 5. The first step in message transmission is the initialization of CAN transmit buffers.

In order to initiate message transmission, the TXBnCTRL.TXREQ bit must be set for each buffer to be transmitted. This can be accomplished by:
1. Writing to the register via the SPI write command
2. Sending the SPI RTS command
3. Setting the TXnRTS pin low for the particular transmit buffer(s) that are to be transmitted

If the message started to transmit but encountered an error condition, the TXBnCTRL.TXERR and CANINTF.MERRF bits will be set and an interrupt will be generated on the INT pin if the CANINTE.MERRE bit is set. If the message is lost, arbitration at the TXBnCTRL.MLOA bit will be set.

The MCU can request to abort a message in a specific message buffer by clearing the associated TXBnCTRL.TXREQ bit. In addition, all pending messages can be requested to be aborted by setting the CANCTRL.ABAT bit. This bit MUST be reset (typically after the TXREQ bits have been verified to be cleared) to continue transmitting messages. The CANCTRL.ABTF flag will only be set if the abort was requested via the CANCTRL.ABAT bit. Aborting a message by resetting the TXREQ bit does NOT cause the ABTF bit to be set.

The CAN message reception is as shown in Figure 6. The RXBnCTRL.RXM bits of the receive buffer set special receive modes. Normally, these bits are cleared to 00 to enable reception of all valid messages as determined by the appropriate acceptance filters. In this case, the determination of whether or not to receive standard or extended messages is determined by the RFXnSIDL.EXIDE bit in the acceptance filter register.

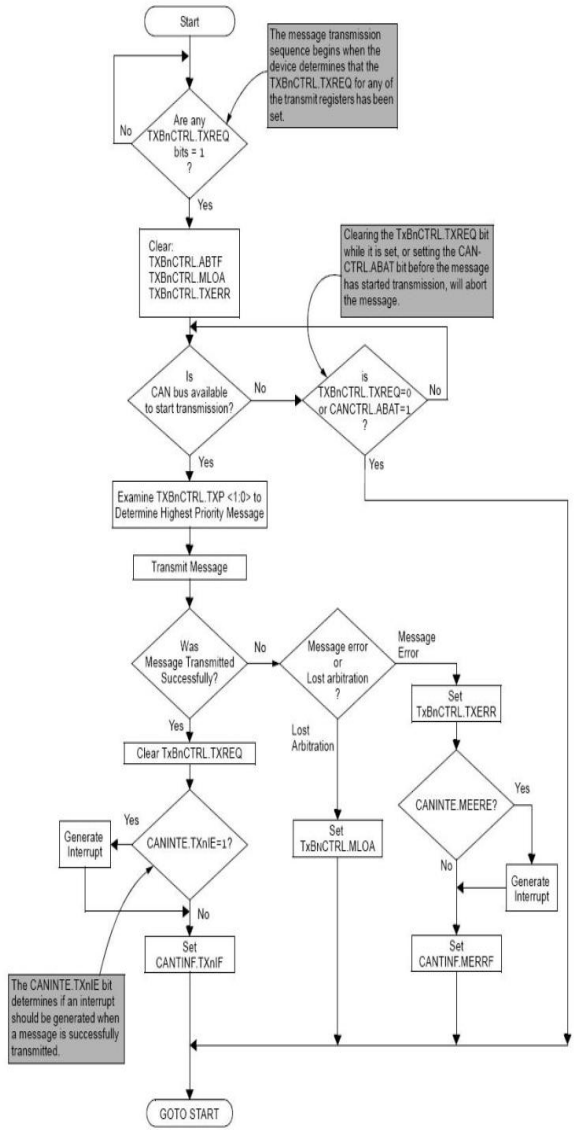

Figure 5. CAN transmission flow

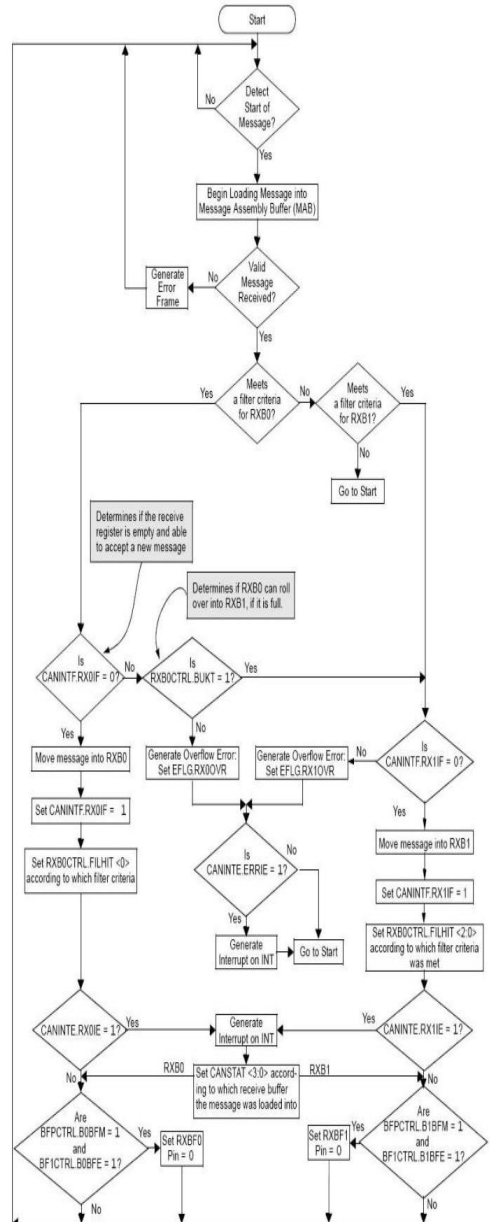

Figure 6. CAN reception flow

in debugging a CAN system and would not be used in an actual system environment.

If the RXBnCTRL.RXM bits are set to 01 or 10, the receiver will only accept messages with standard or extended identifiers, respectively. If an acceptance filter has the RFXnSIDL.EXIDE bit set such that it does not correspond with the RXBnCTRL.RXM mode, that acceptance filter is rendered useless. These two modes of RXBnCTRL.RXM bits can be used in systems where it is known that only standard or extended messages will be on the bus.

If the RXBnCTRL.RXM bits are set to 11, the buffer will receive all messages, regardless of the values of the acceptance filters. Also, if a message has an error before the EOF, that portion of the message assembled in the MAB before the error frame will be loaded into the buffer. This mode has some value

## VII. CONCLUSION

In this paper, the CAN-bus based communication system for accident avoidance system is designed. System can be upgraded easily and use of CAN reduces wiring to a great extent. Real-time, reliability and flexibility, all these characteristics make CAN BUS an indispensable network communication technology applied in automobile network communication field. Also, use of ARM 7 processor ensures fast operation, high efficiency, low cost, low power and higher performance.

## ACKNOWLEDGEMENT

I Kashyap M. Joshi would like to thank everyone, including: parents, teachers, family, friends, and in essence, all sentient beings for their help and support this paper would not have been possible.Especially,I dedicate my acknowledgment of gratitude toward my mentor and Co-author Prof. Vipul J. Gohil for his guidance and support.

## REFERENCES

[1] Li Ran, Wu Junfeng, Wang Haiying, Li Gechen. "Design Method of CAN BUS Network Communication Structure for Electric Vehicle", IFOST 2010 Proceedings IEEE.

[2] Yujia Wang, Hao Su, Mingjun Zhang., "CAN-Bus-Based Communication System Research for Modular underwater Vehicle", 2011 IEEE DOI 10.1109/ICICTA.2011.

[3] Chin E. Lin, S. F. Tai, H. T. Lin, T. P. Chen, P. K. Chang, C. C. Kao "Prototype Of A Small Aircraft Avionics Using Hybrid Data Bus Technology" 2005 IEEE

[4] Chin E. Lin, Hung-Ming Yen, "A Prototype Dual Can-Bus Avionics System For Small Aircraft Transportation System" 2006 IEEE.

[5] V. Claesson, C. Ekelin, N. Suri, "The event-triggered and time- triggered medium-access methods", 6th IEEE International Symposium on Object-Oriented Real-Time Distributed computing, May 14-16, 2003, pp. 131-134.

[6] Chris Quigley, Richard McLaughlin, "Electronic System Inte-Gration For Hybrid And Electric Vehicles".

[7] Chin E. Lin, H. M. Yen, "Reliability And Stability Survey On Can-Based Avionics Network For Small Aircraft" -7803-9307- 4/05/2005 IEEE.

[8] Feng Xia, Xiaohua Dai, Zhi Wang, and Youxian Sun, "Feedback Based Network Scheduling of Networked Control Systems"2005 IEEE.

[9] Ashwini S. Shinde, Prof. Vidhyadhar B. Dharmadhikari, "Controller Area Network for Vehicle Automation" International Journal of Emerging Technology and Advanced Engineering www.ijetae.com ISSN 2250-2459, Volume 2, Issue 2, February 2012.

[10] Mazran Esro, Amat Amir Basari, Siva Kumar S, A. Sadhiqin M I, Zulkifli Syariff, "Controller Area Network (CAN) Application in Security System" World Academy of Science, Engineering and Technology 35 2009.